\documentclass[twoside]{ae100prg}
\newcommand{\be}{\begin{equation}}
\newcommand{\ee}{\end{equation}}

\usepackage{graphicx}
\usepackage[breaklinks]{hyperref}
\usepackage{booktabs}

\begin{document}
\title{Higher-dimensional black holes}

\author{Harvey S. Reall}

\address{Department of Applied Mathematics and Theoretical Physics, University of Cambridge, Wilberforce Road, Cambridge CB3 0WA, United Kingdom}

\email{hsr1000@cam.ac.uk}

\begin{abstract}
This article reviews black hole solutions of higher-dimensional General Relativity. The focus is on stationary vacuum solutions and recent work on instabilities of such solutions.
\end{abstract}

\section{Introduction}

General Relativity (GR) in $D>4$ spacetime dimensions has been actively investigated for more than a decade. There are several reasons for this interest in higher dimensions, and higher-dimensional black holes in particular.

\begin{enumerate}
\item
Statistical calculation of black hole entropy using string theory. This was first done for certain $D=5$ black holes \cite{Strominger:1996sh}. Each entropy calculation is a check on the theory, irrespective of the dimension. Hence the study of higher-dimensional black holes is a worthwhile contribution to developing a theory of quantum gravity.
\item
The gauge/gravity correspondence \cite{Maldacena:1997re} relates the properties of black holes in $D$ dimensions to strongly coupled, finite temperature, quantum field theory in $D-1$ dimensions. This provides a way of calculating certain field theory quantities which are very hard to determine by more traditional methods.
\item
The possibility of producing tiny higher-dimensional black holes at colliders in certain "brane-world" scenarios \cite{Kanti:2008eq}.
\item
Higher-dimensional black hole spacetimes might have useful mathematical properties. For example, analytically continued versions of black hole solutions have been used to obtain explicit metrics on compact Sasaki-Einstein spaces \cite{Cvetic:2005vk}.
\item
An explicit higher-dimensional solution might provide a clean example of some important effect in GR. A nice example of this is the frame-dragging effect exhibited by the "black Saturn" solution (see below). 
\item
Progress in quantum field theory has been made by considering $D$ different from $4$, and fields different from those of the Standard Model. In the same spirit, perhaps we will arrive at a better understanding of GR by allowing the parameter $D$ to take values other than $4$  \cite{Emparan:2008eg}. 
\end{enumerate}

This article is a brief, selective, review of higher-dimensional black hole solutions. The scope is limited to solutions of the vacuum Einstein equation without cosmological constant. I shall discuss two types of black hole. (i) Black hole solutions of Kaluza-Klein theory that are static in the higher-dimensional sense. This includes black string solutions. (ii) Asymptotically flat black hole solutions. In each case, there has been recent progress in demonstrating the existence of instabilities of certain solutions and so special attention is given to this topic. 

We close this introduction by presenting the simplest higher-dimensional black hole, the $D$-dimensional Schwarzschild solution:
\be
 ds^2=-fdt^2 + \frac{dr^2}{f} + r^2 d\Omega^2_{D-2}, \qquad f = 1 - \left( \frac{r_+}{r} \right)^{D-3}
\ee
where $d\Omega_{D-2}^2$ is the line-element on a unit round $S^{D-2}$ and the event horizon is at $r=r_+$. 
\section{Black holes in Kaluza-Klein theory}

Consider vacuum GR with a compact Kaluza-Klein circle. We are interested in black hole solutions of this theory which are asymptotically flat in a Kaluza-Klein sense, which means that, at large distance in the non-compact directions, the metric approaches that of Minkowski space with a compact circle of circumference $L$:
\be
  ds^2 \approx -dt^2 + dr^2 + r^2 d\Omega_{D-2}^2 + dz^2, \qquad z \sim z + L
\ee
I shall describe the known {\it static} black hole solutions of this theory. For a more detailed recent review see Ref. \cite{Horowitz:2011cq}.

The simplest such black hole solution is the product of the $(D-1)$-dimensional Schwarzschild solution with a circle of circumference $L$. This gives a {\it black string} solution for which cross-sections of the event horizon have topology $S^1 \times S^{D-3}$. In the decompactified limit $L \rightarrow \infty$ it gives a black string of infinite length.

Another solution of this theory describes a black hole of topology $S^{D-2}$ localized on the Kaluza-Klein circle. Such solutions are not known explicitly. Solutions describing black holes with radius much smaller than $L$ have been constructed perturbatively \cite{Harmark:2003yz,Gorbonos:2004uc}. Larger black hole solutions have been constructed numerically for $D=5,6$ \cite{Sorkin:2003ka,Kudoh:2004hs}. They cannot become arbitrarily large: there is an upper bound on their mass determined by $L$. In fact, as one moves along this family of solutions, starting from a small black hole, the mass increases to a maximum and then decreases \cite{Kudoh:2004hs,Headrick:2009pv}.

The higher-dimensional Schwarzschild solution is stable against linearized gravitational perturbations \cite{Ishibashi:2003ap}. However, black strings suffer from the Gregory-Laflamme (GL) instability \cite{Gregory:1993vy}. If $r_+/L$ is less than a certain $D$-dependent critical value then there exist linearized gravitational perturbations which grow exponentially with time. These perturbations break the translational symmetry around the KK circle. If $r_+/L$ exceeds the critical value then the string  is believed to be stable. So "thin" strings are unstable and "fat" strings are stable. Taking the limit $L \rightarrow \infty$ shows that uncompactified black strings are unstable.

Since its discovery, the nonlinear evolution of this instability has been the subject of considerable debate. It was first proposed that the black string would "pinch off" to form a localized KK black hole or holes \cite{Gregory:1993vy}. This cannot happen classically owing to the result that black holes cannot bifurcate. Instead, a singularity would have to form at the horizon at the moment of the pinching off, but perhaps this is resolved by quantum effects. 

Recent numerical simulations (with $D=5$) support this picture \cite{Lehner:2010pn}. These start from a perturbation which is sinusoidal along the flat direction. When evolved, the perturbation becomes more inhomogeneous, with the configuration reaching a transient state resembling a line of localized black holes connected by thin threads of black string. However, these threads are in turn unstable and suffer the Gregory-Laflamme instability, but on a shorter timescale, leading to smaller black holes connected by even thinner strings. The process appears to continue on smaller and smaller scales, in a self-similar manner, but in a finite total time as measured by an observer far from the string. The curvature of (parts of) the horizon becomes large in this process, so it seems that a naked singularity does indeed form. This is strong evidence against the validity of the cosmic censorship hypothesis in higher dimensions. 

The black strings described above are invariant under translations around the KK circle. For this reason they are called {\it uniform} black strings. Some time ago, it was conjectured that there should exist a 1-parameter family of static {\it nonuniform} black strings which lack this translational symmetry, and bifurcate from the uniform black string family at the critical value of $r_+/L$ discussed above \cite{Horowitz:2001cz}. Such solutions were subsequently constructed perturbatively, for infinitesimal non-uniformity \cite{Gubser:2001ac}. Fully nonlinear solutions have been constructed numerically \cite{Wiseman:2002zc,Kleihaus:2006ee,Sorkin:2006wp,Figueras:2012xj}. Just like the localized KK black holes, there is an upper bound on their mass determined by $L$.

Ref. \cite{Kol:2002xz} conjectured that the families of localized KK black hole solutions and non-uniform black strings should merge at a common limiting solution. The idea is that, moving along the family of localized black holes,  they increase in size until they fill the KK circle, and then transition to a black string of high non-uniformity. Numerical evidence supporting this suggestion was obtained in Refs.  \cite{Sorkin:2003ka,Kudoh:2004hs,Headrick:2009pv}.

The perturbative construction of non-uniform black strings reveals that, for $D \le 13$, infinitesimally non-uniform black strings have lower horizon area than a uniform black string of the same mass, whereas for $D>13$, they have greater horizon area \cite{Sorkin:2004qq}. This suggests that infinitesimally non-uniform black strings should be classically unstable for $D \le 13$ but stable for $D > 13$. It also suggests that a non-uniform black string could be an endpoint of the GL instability for $D > 13$.

Very recently, Ref. \cite{Figueras:2012xj} has performed the first study of the stability of non-uniform black strings with finite non-uniformity. The results confirm the perturbative results for infinitesimal non-uniformity. It was found that the instability persists to large non-uniformity for $D \le 11$. Strong evidence that non-uniform strings with $D> 13$ are all stable was presented. Solutions with $D=12,13$ were constructed for the first time in Ref. \cite{Figueras:2012xj}. It was found that there is a maximum mass as one moves along the family of non-uniform black strings. Solutions before the maximum are unstable whereas solutions after the maximum appear to be stable, and can have greater horizon area than a uniform string of the same mass.

Properties of these different solutions can be summarized in a plot of horizon area against mass, for fixed $L$. This is useful in understanding possible time evolution of instabilities of the various solutions. Horizon area increases, and energy decreases (via emission of gravitational waves) so the final state of an instability must lie "up and left" of the initial state on such a diagram. 

\begin{figure}[h!]
\begin{center}
\includegraphics[scale=0.4]{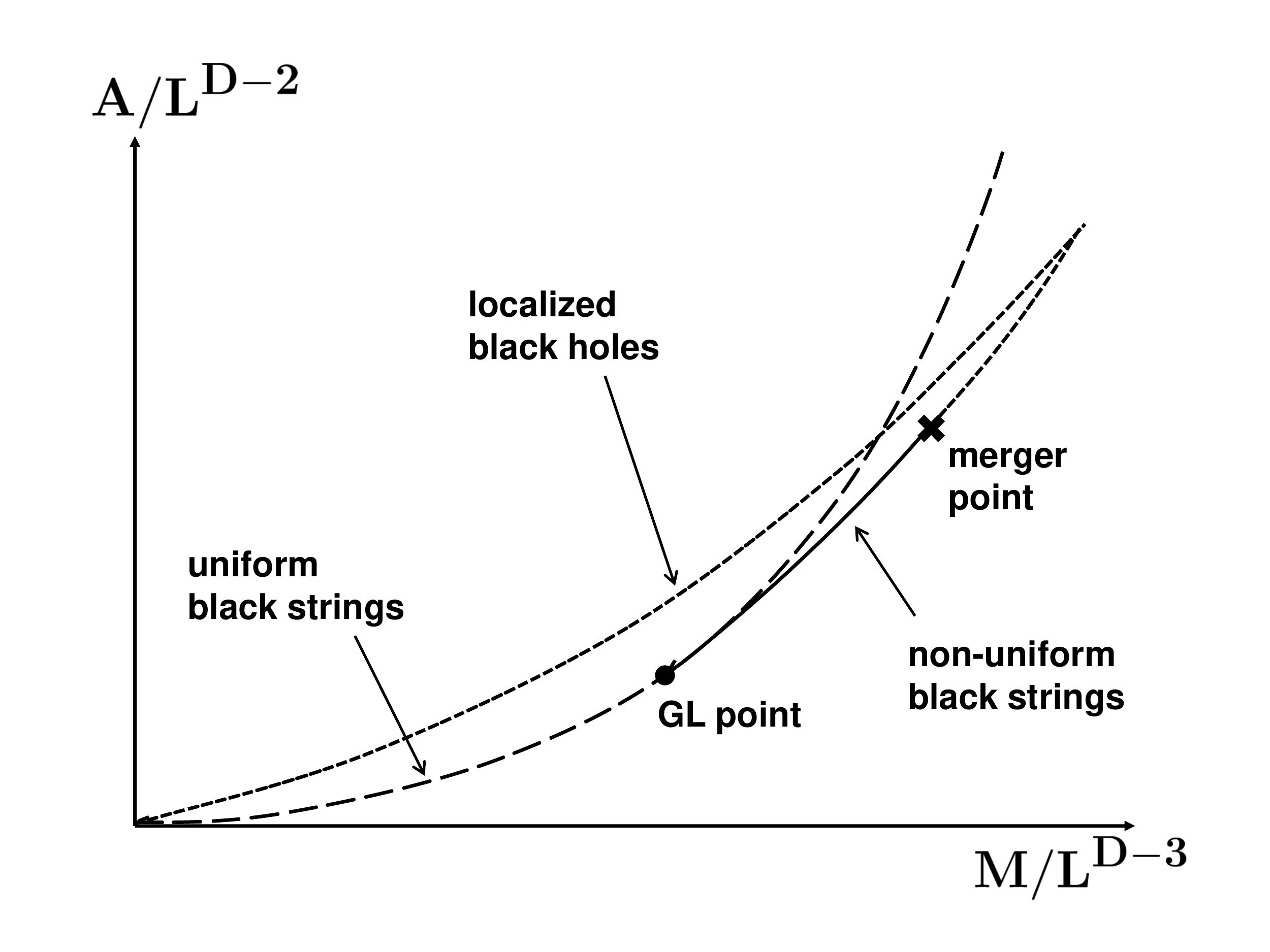}
\end{center}
\caption{Schematic plot of horizon area against mass for Kaluza-Klein black holes/strings with $D=5,6$ based on results of Refs \cite{Kudoh:2004hs,Headrick:2009pv}. The dashed curve is the uniform black string branch, the solid curve the non-uniform string branch and the dotted curve the localized black hole branch.  It seems likely that this will be the qualitative behaviour for all $D \le 11$. (Plot reproduced from Ref. \cite{Figueras:2012xj}.)
}
 \label{figure:phasediagram1a}
\end{figure}

Fig. \ref{figure:phasediagram1a} shows the qualitative form of a plot of horizon area against mass for the cases $D=5,6$ (see Ref. \cite{Horowitz:2011cq} for a more quantitative plot). Recall that there is a maximum mass solution along the localized black hole branch of solutions. By the first law, an extremum of the mass must also be an extremum of horizon area, and this results in the cusp shown in the figure. 

The stability of localized black holes has not been studied. This could be done using the method of Ref. \cite{Figueras:2012xj}. This method shows that an instability appears as one moves through a maximum of mass along a branch of solutions. Since small localized black holes are expected to be stable, it seems likely that, moving along the branch of solutions, they will be stable until the maximum is reached and unstable thereafter. Hence, in Fig. \ref{figure:phasediagram1a}, unstable solutions lie on the part of the uniform string curve extending from the origin to the GL point, along the non-uniform string curve from the GL point to the merger point, and then along the localized black hole curve from the merger point to the cusp. 

\begin{figure}[h!]
\begin{center}
\includegraphics[scale=0.4]{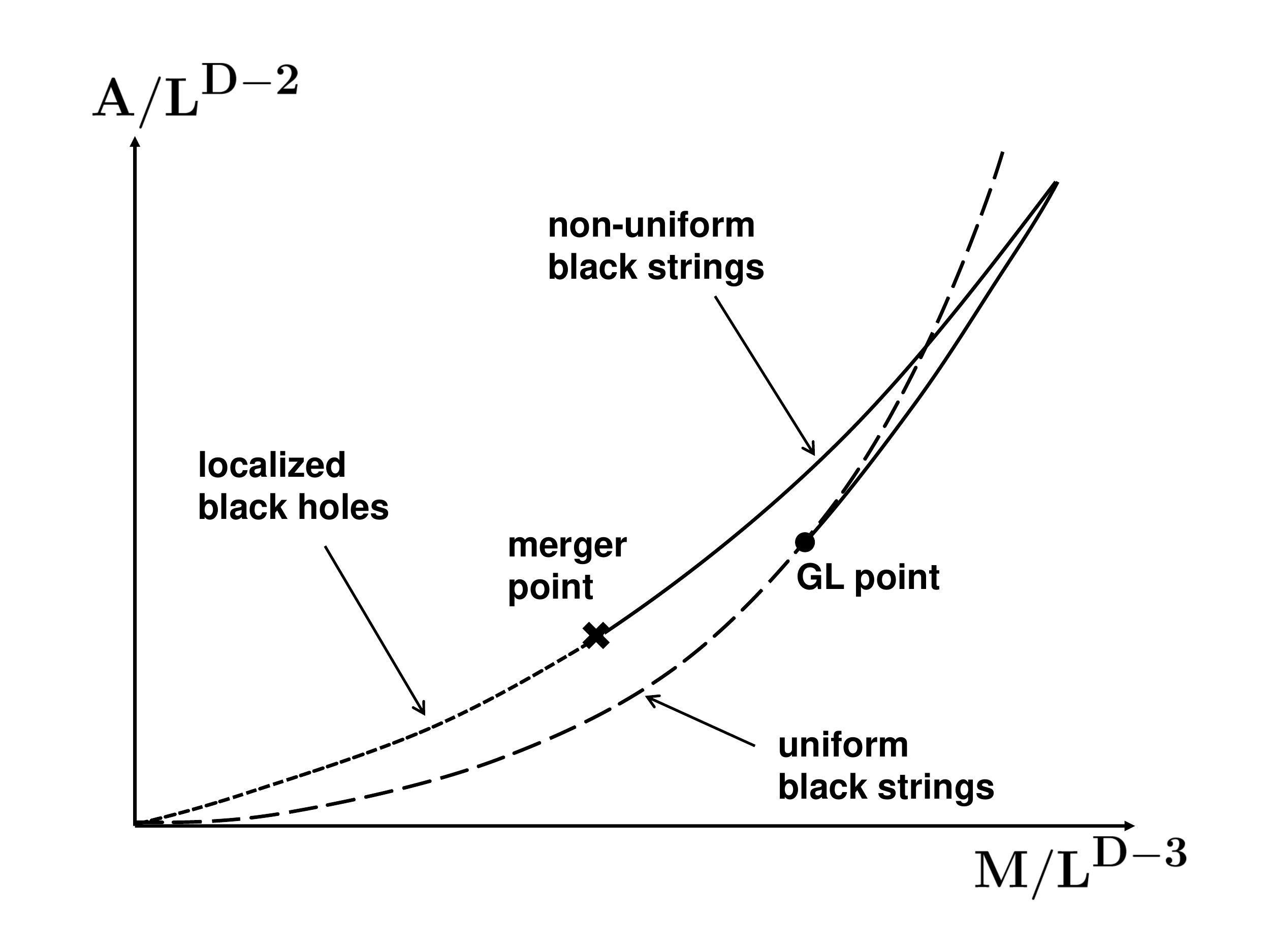}
\end{center}
\caption{Schematic plot of horizon area against mass for Kaluza-Klein black holes/strings with $D=12,13$ based on results of Ref. \cite{Figueras:2012xj}. The localized black hole curve is conjectural. (Plot reproduced from Ref. \cite{Figueras:2012xj}.)}
 \label{figure:phasediagram1}
\end{figure}

For $D=12,13$, the change in behaviour of the non-uniform string branch implies that the diagram must change to that of Fig. \ref{figure:phasediagram1} where the cusp now appears along the non-uniform black string branch. (Note that localized black hole solutions have not been constructed for $D>6$ so this part of the Figure is conjectural.) In this case, unstable solutions lie on the part of the uniform string curve extending from the origin to the GL point and along the non-uniform string curve from the GL point to the cusp. 

The results of Ref. \cite{Figueras:2012xj} suggest that there is no maximum mass non-uniform black string for $D=11$. Hence it seems likely that Fig. \ref{figure:phasediagram1a} gives the behaviour for all $D \le 11$.

\begin{figure}[h!]
\begin{center}
\includegraphics[scale=0.4]{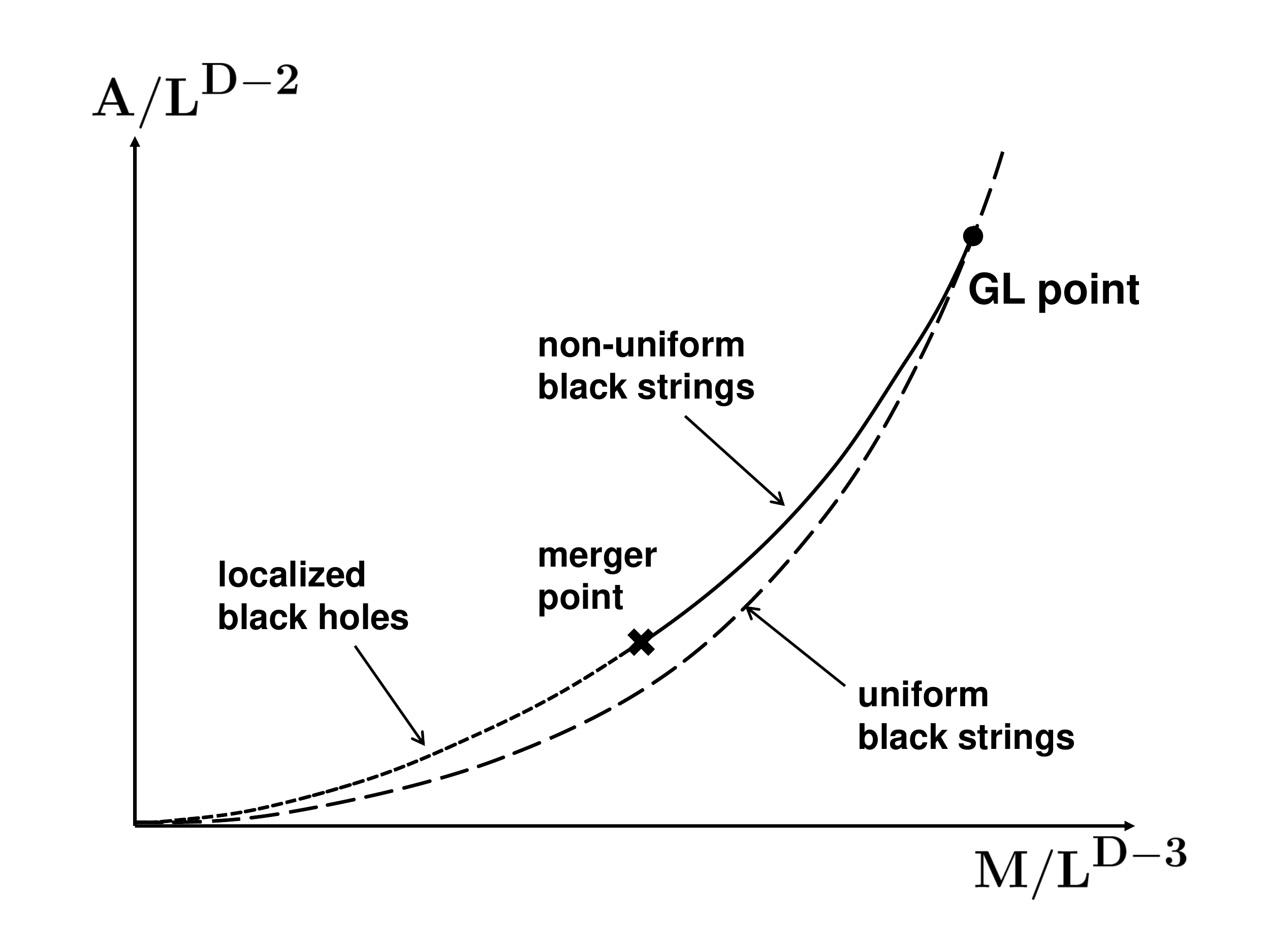}
\end{center}
\caption{Schematic plot of horizon area against mass for Kaluza-Klein black holes/strings with $D>13$ based on results of Ref. \cite{Figueras:2012xj}. The localized black hole curve is conjectural. (Plot reproduced from Ref. \cite{Figueras:2012xj}.)}
\label{figure:phasediagram2}
\end{figure}

For $D>13$, the results for non-uniform black strings, and the simplest guess for the behaviour of localized black holes, results in the phase diagram shown in Fig. \ref{figure:phasediagram2}. Unstable solutions lie on the part of the uniform string curve extending from the origin to the GL point.

\section{Asymptotically flat black holes}

\subsection{Introduction}

For a comprehensive 2008 review of higher-dimensional black holes, see Ref. \cite{Emparan:2008eg}. For more recent reviews, see Refs. \cite{horowitz,japanreview}.

Recall that angular momentum is defined in terms of an antisymmetric matrix $J_{ij}$ where $i,j$ run over the $D-1$ spatial dimensions (e.g. for a particle, $J_{ij} = x_i p_j - x_j p_i$). For $D=4$, $J_{ij}$ is equivalent to a vector $J_i$ and so, by choosing $z$-axis aligned with this vector, one can write $J_{ij}$ in terms of a single component $J$. For $D>4$ dimensions, the best one can achieve by a choice of axes is a block-diagonal form for $J_{ij}$ where each block is a $2 \times 2$ antisymmetric matrix specified by a component $J_I$, where $I=1, \ldots, N=[(D-1)/2]$. 

There are two families of explicit black hole solutions of the vacuum Einstein equation in $D>4$ spacetime dimensions: Myers-Perry black holes \cite{Myers:1986un} and black rings \cite{Emparan:2001wn,Pomeransky:2006bd}.

\subsection{Myers-Perry black holes}

The Myers-Perry solution is the generalization of the Kerr solution to $D$ spacetime dimensions. See Ref. \cite{Myers:2011yc} for a more detailed review. Myers-Perry black holes share many properties with the Kerr solution. Cross-sections of the event horizon have spherical topology $S^{D-2}$. The solution is uniquely parameterized by its mass $M$ and its angular momenta $J_I$. The general Myers-Perry solution has $N=[(D-1)/2]$ commuting rotational Killing vector fields $\partial/\partial \phi_I$. The Myers-Perry solution with all $J_I=0$ reduces to the $D$-dimensional Schwarzschild solution.

\begin{figure}[h!]
\begin{center}
\includegraphics[scale=1.0]{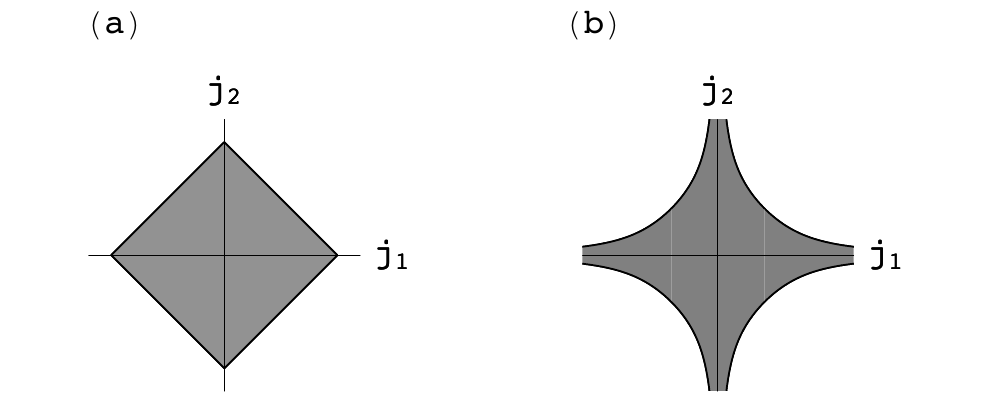}
\end{center}
\caption{Parameter space of Myers-Perry black holes for (a) $D=5$ and (b) $D=6$. The axes are dimensionless angular moment $j_I \sim J_I M^{-(D-2)/(D-3)}$. Non-extreme black holes correspond to the shaded region. The boundary of this region corresponds to extreme black holes, except for the vertices of the square, which describe singular solutions. (Plot reproduced from Ref. \cite{Emparan:2008eg}.)}
 \label{figure:MPphasespace}
\end{figure}

Recall that, for given $M$, the Kerr solution has an upper bound on its angular momentum $|J| \le M^2$ and saturating this bound gives the extreme Kerr solution with a regular, but degenerate horizon. For $D=5$, there is a similar upper bound on the angular momenta of the MP solution: for given $M$, regular black holes have $J_1,J_2$ confined to a square region centred on the origin in the $(J_1,J_2)$ plane, see Figure. \ref{figure:MPphasespace}.
Saturating this bound gives a black hole with a degenerate horizon except when one of the angular momenta vanishes (the vertices of the square), in which case the spacetime is singular with no horizon.

For $D>5$, there is a qualitative difference between MP and Kerr. It is possible for the angular momenta to be arbitrarily large for fixed $M$. A "singly spinning" black hole, i.e., one with $J_2 = J_3 = \ldots = J_N = 0$, has no upper bound on $J_1$. More generally, it is possible for some of the angular momenta to be very large if others are small. See Figure \ref{figure:MPphasespace} for the $D=6$ case. Ref. \cite{Emparan:2003sy} studied the geometry of singly spinning MP black holes in the "ultraspinning" limit of very large rotation. It was found that the black hole becomes flattened into the plane of rotation, so that it resembles a rotating pancake. The geometry near the intersection of the axis of rotation with the horizon approaches that of a black {\it membrane}: the product of a $(D-2)$-dimensional Schwarzschild solution with two flat directions. Black membranes suffer from the Gregory-Laflamme instability. Hence it was conjectured in Ref. \cite{Emparan:2003sy} that rapidly rotating $D>5$ MP black holes are classically unstable.

Confirmation of this conjecture required a study of linearized gravitational perturbations of MP solutions. For $D=4$, the study of gravitational perturbations of a Kerr black hole is simplified by the remarkable "decoupling" phenomenon discovered by Teukolsky \cite{Teukolsky:1972my}, which reduces the problem to a PDE for a single scalar quantity. Unfortunately, decoupling does not occur for $D>4$ \cite{Durkee:2010qu} and so one has to solve a large set of coupled PDEs instead. This was done in Ref. \cite{Dias:2009iu}, which studied a class of linearized perturbations of a singly spinning MP solution, restricting to perturbations that preserve the symmetries of the MP solution, i.e., stationarity and the rotational symmetries. For fixed $M$, it was found (numerically) that there exists a critical value of $J_1$ for which a singly spinning $D>5$ MP solution admits a non-trivial stationary linearized gravitational perturbation. This was interpreted as the "threshold mode" indicating the onset of the ultraspinning instability of Ref. \cite{Emparan:2003sy}, i.e., black holes with larger $J_1$ should be unstable.

There is a more symmetrical class of MP solutions: those with odd $D$ and $J_1 = J_2 = \ldots = J_N \equiv J$. Such solutions are "cohomogeneity-1": they depend non-trivially only on the radial coordinate. This implies that the equations governing gravitational perturbations are ODEs rather than PDEs \cite{Kunduri:2006qa}. However, in this case, there is an upper (extremality) bound on $J$ for given $M$, i.e., there is no reason to expect an ultraspinning instability. The $D=5$ case was studied in Ref. \cite{Murata:2008yx} and no evidence of any instability was found. However, Ref. \cite{Dias:2010eu} showed that, for $D=9$, with $J$ close to the upper bound, there are linearized gravitational perturbations which grow exponentially with time, i.e., an instability. This was extended to $D=7$ in Ref. \cite{Dias:2011jg}, which also considered a class of MP solutions interpolating between singly spinning and cohomogeneity-1 and determined the threshold of instability in this case.

Although decoupling of perturbations does not occur for higher-dimensional black holes, it does occur for the {\it near-horizon geometry} of an extreme vacuum black hole \cite{Durkee:2010qu}. Ref. \cite{Durkee:2010ea} used this to argue that the instability of near-extreme cohomogeneity-1 MP solutions can be predicted analytically, thereby extending the result to any odd $D>5$.  Ref. \cite{Tanahashi:2012si} used the same approach to show that MP solutions with even $D$ and $J_1 = J_2 = \ldots = J_N$ (which are cohomogeneity-2) also are unstable near extremality. 

The perturbations discussed so far are invariant under the Killing vector field $\Omega_I \partial/\partial \phi_I$ where $\Omega_I$ are the angular velocities of the horizon.\footnote{
The exception is the analysis in Ref.  \cite{Murata:2008yx} of the cohomogeneity-1 $D=5$ case, i.e., $J_1 = J_2$. The results are consistent with stability of this solution.}  This makes the resulting equations easier to solve. In the singly spinning case, this means that the perturbations are axisymmetric. {\it Non}axisymmetric perturbations of singly spinning MP have been studied using full-blown numerical relativity \cite{Shibata:2009ad,Shibata:2010wz}.

Ref. \cite{Shibata:2009ad} studied the $D=5$ case and found that, for $J_1$ near to the upper bound, an initially small non-axisymmetric perturbation grows in amplitude. It was not possible to evolve the system long enough to determine the endpoint of this instability. The corresponding problem for $D=6,7,8$ was studied in Ref. \cite{Shibata:2010wz}. An instability was found for large enough dimensionless angular momentum $j_1 \equiv J_1M^{-(D-2)/(D-3)}$. This instability appears at a lower value of $j_1$ than the axisymmetric instability discussed above, i.e., the nonaxisymmetric instability is the first one to appear as the angular momentum is increased. In this case, it was possible to follow the long time evolution of the instability. It was found that the perturbed black hole emits gravitional waves, which carry away angular momentum (and energy), and the black hole finally settles down to a (presumably stable) Myers-Perry black hole with a lower value of $j_1$. It is not clear whether this also happens for $D=5$ or whether the evolution of the instability is qualitatively different in this case.

Finally we should note that there is an instability that afflicts {\it extreme} black holes, including extreme Myers-Perry. Consider a massless scalar field in the background of a Kerr black hole. In the non-extreme case, it has been proved that, for any initial data (decaying at infinity), the scalar field and all its derivatives decay on, and outside, the horizon \cite{Dafermos:2010hd}. However, the extreme case is qualitatively different. For axisymmetric initial data, the scalar field decays on, and outside, the event horizon \cite{Aretakis:2011gz}. However, a transverse derivative of the scalar field at the horizon generically does not decay (hence the energy-momentum tensor does not decay), and higher transverse derivatives at the horizon blow up, i.e., they become large at late time \cite{Aretakis:2012ei}. Therefore the scalar field is stable in the background of an arbitrarily non-extreme black hole, but unstable in the background of an exactly extreme black hole. Note that the instability involves power-law, rather than exponential, growth in time.

The scalar field here should be regarded as a toy model for linearized gravitational perturbations, which suggests that an extreme Kerr black hole should suffer a gravitational instability. Ref. \cite{Lucietti:2012sf} showed that this is indeed the case. This reference also showed that similar non-decay and blow-up results hold for a massless scalar field in {\it any} extreme black hole spacetime. Hence extreme black holes generically are unstable, as conjectured in Ref. \cite{Marolf:2010nd}.

\subsection{Black rings}

A black ring is an asymptotically flat black hole for which horizon cross-sections have topology $S^1 \times S^{D-3}$. There is a heuristic argument for the existence of such objects in vacuum gravity. Take a finite segment of (uniform) black string and imagine forming it into a loop. The loop would collapse under its own gravity and tension. However, if it rotates then Newtonian arguments suggest that the resulting centrifugal repulsion can balance the gravitational and tension forces, resulting in a stationary black ring. 

\begin{figure}[h!]
\begin{center}
\includegraphics[scale=0.7]{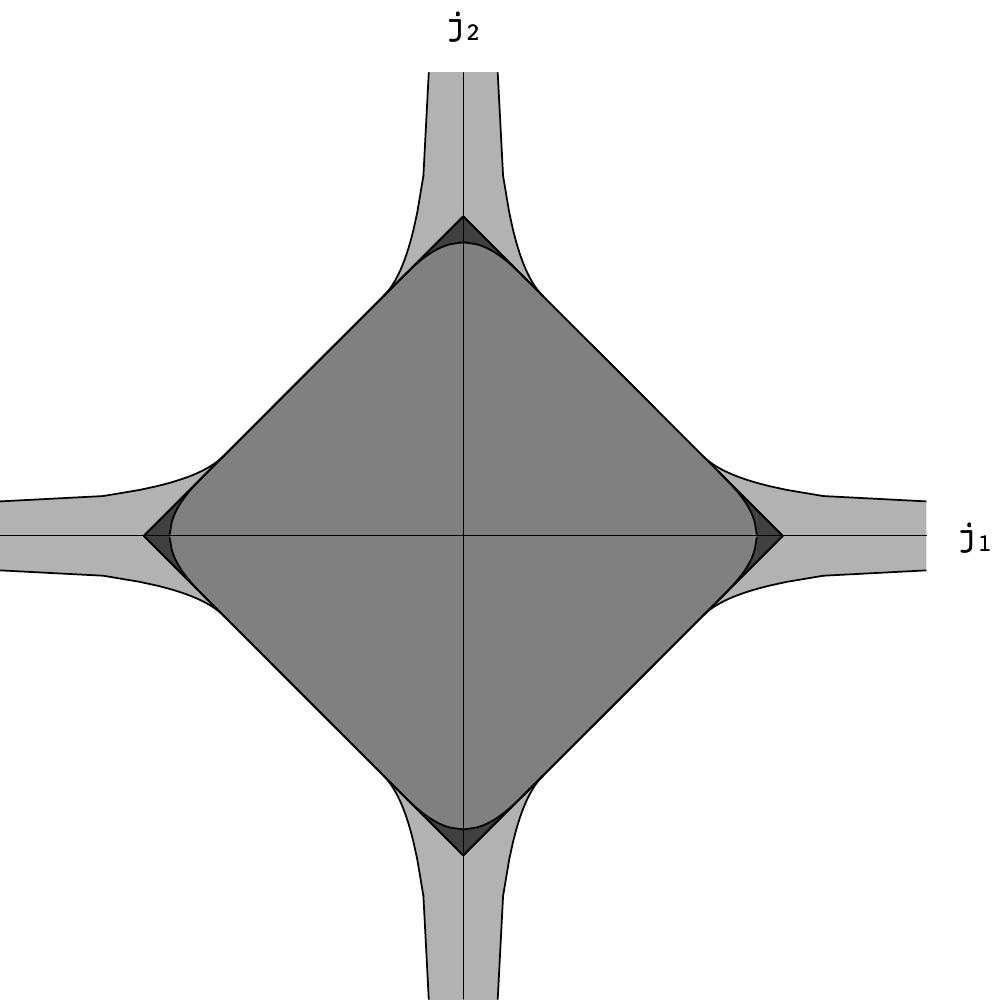}
\end{center}
\caption{Phase space of $D=5$ Myers-Perry black holes and black rings. See Fig. \ref{figure:MPphasespace} for notation. For each point of the light grey regions there exists a "thin" black ring. For each point of the mid-grey region there exists a MP black hole. For each point of the dark grey region there exists a MP black hole, a fat black ring and a thin black ring. (Plot reproduced from Ref. \cite{Emparan:2008eg}.)}
 \label{figure:MPring}
\end{figure}

This heuristic argument for the existence of black rings is confirmed by explicit solutions, which are known only for the special case $D=5$ \cite{Emparan:2001wn,Pomeransky:2006bd}. These solutions are the first examples of asymptotically flat black holes with horizon cross-sections of non-spherical topology. They form a 3-parameter family and have 2 commuting rotational symmetries. Unlike the Myers-Perry solutions, black rings are {\it not} uniquely labelled by $M$ and $J_I$: it is possible for there to be two black rings with the same values for these quantities. Furthermore, a black ring can have the same value for $M,J_I$ as a Myers-Perry black hole. See Fig. \ref{figure:MPring}. Hence the existence of black rings shows that black hole uniqueness cannot be straightforwardly extended to higher dimensions. 

An important difference between black rings and $D=5$ MP solutions is that, for given $M$, there is a {\it lower} bound on the angular momentum $J_1$, and (for $J_2=0$) no upper bound. Black rings do not admit a regular static limit, as expected from the heuristic argument for their existence. 

It is convenient to divide black rings into two subclasses according to the sign of the heat capacity at constant angular momenta $c_J$. Black rings with $c_J<0$ are called "thin" and those with $c_J>0$ are called "fat". The terminology arises from the geometry of the horizon. Thin rings look more like hula-hoops and fat rings more like bagels. Rings of each type are uniquely labelled by $M,J_I$. 

Heuristic arguments indicate that fat black rings probably are unstable. Ref. \cite{Arcioni:2004ww} used the Poincar\'e turning point method to argue that fat black rings with small $c_J$ should have one more "unstable mode" than thin rings. Ref. \cite{Elvang:2006dd} considered certain singular deformations of the black ring solution to determine an effective potential for radial deformations of the ring. It was found that thin rings sit at a local minimum of the potential but fat rings correspond to a local maximum, suggesting instability. 

These results were confirmed by the analysis of Ref. \cite{Figueras:2011he}, which introduced a new method for studying black hole stability. If a black hole is stable then any small perturbation of it must eventually settle down to a black hole belonging to the same family, with a small change in its parameters. If one restricts to rotationally symmetric perturbations, so that angular momentum is conserved, then, by using increase of horizon area and decrease of (Bondi) energy, one can deduce that the initial data describing the perturbed black hole must satisfy a certain inequality relating its mass, angular momenta and horizon area. If one can find an initial perturbation that violates this inequality then the black hole cannot be stable. The nice thing about this method is that it requires only the construction of initial data, rather than determining the full time-evolution of the perturbation. Using this approach, it was shown that fat black rings suffer from a rotationally symmetric instability.

Rings with large $J_1$ (for given $M$) are very thin. As $J_1 \rightarrow \infty$, the geometry near a section of the ring approaches that of a boosted uniform black string. Since the latter suffers from the Gregory-Laflamme instability, it seems very likely that black rings with large $J_1$ will be classically unstable. This instability breaks rotational symmetry so it cannot be studied using the approach of Ref. \cite{Figueras:2011he}. Demonstrating the existence of this instability will require a study of linearized gravitational perturbations of black rings (or full-blown numerical GR). This has not yet been attempted.  

In summary, fat black rings are known to be unstable and very thin black rings are believed to be unstable. But it is not known whether all black rings are unstable or whether some thin, but not too thin, rings are stable.

So far we have been discussing $D=5$ black rings. Explicit black ring solutions are not known for $D>5$. However, approximate solutions describing very thin black rings have been constructed using the perturbative "blackfold" approach to be described below \cite{Emparan:2007wm}. Recently, Ref. \cite{Kleihaus:2012xh} reported a breakthrough in determining $D>5$ dimensional black ring solutions by numerical solution of the Einstein equation. Results were presented for black rings with $D=6,7$ with a single non-vanishing angular momentum $J_1$. Their properties appear similar to those of $D=5$ black rings, and agree with the predictions of the blackfold approach when the radius becomes large.

\subsection{Black Saturn and generalizations}

Given the existence of black rings and Myers-Perry black holes, it is natural to ask whether one can "superpose" these solutions to construct a "Black Saturn" describing a MP black hole with a concentric black ring. Of course the Einstein equation is nonlinear so this is highly non-trivial. Nevertheless, solution generating techniques have been used to construct such a solution with $D=5$ \cite{Elvang:2007rd}. 

Black Saturn is the first example of an explicit stationary, asymptotically flat, vacuum, regular, multi-black hole solution. It provides an interesting demonstration of the frame-dragging effect: if one sets the (Komar) angular momentum of the MP black hole to zero then its angular velocity is non-zero because the horizon generators are dragged around by the rotation of the black ring.  

Solution generating techniques have also been used to construct solutions with multiple concentric black rings. For example the di-ring of Refs.  \cite{Iguchi:2007is,Evslin:2007fv} describes a pair of concentric black rings lying in the same plane. Ref. \cite{Elvang:2007hs} gave a solution describing a pair of concentric black rings lying in orthogonal planes.

\subsection{Classification of asymptotically flat black holes}

For $D=4$, the black hole uniqueness theorem provides a complete classification of stationary vacuum black holes. For $D>4$, the known solutions show that the situation is much more complicated and the uniqueness theorem does not generalize in a simple way. However, it is useful to explore whether some aspects of the $D=4$ theorem can be generalized to $D>4$.

Uniqueness of static vacuum black holes turns out to generalize straightforwardly to $D>4$: the only asymptotically flat static vacuum black holes solution is the Schwarzschild solution, for any $D \ge 4$ \cite{Gibbons:2002av}.
 
For non-static black holes, the first logical step in the $D=4$ uniqueness theorem is Hawking's topology theorem \cite{Hawking:1973uf}, stating that horizon cross-sections must have $S^2$ topology. This result has been generalized to $D>4$ dimensions, with the result that horizon cross-sections must admit a metric of positive scalar curvature \cite{Galloway:2005mf}. This is a topological restriction on the horizon. For $D=4$ it reduces to Hawking's result. For $D=5$ it implies that the horizon cross-section must be either $S^3$ (or a quotient), $S^1 \times S^2$, or a connected sum of these. It is striking that $S^3$ and $S^1 \times S^2$ are precisely the topologies realized by the known Myers-Perry and black ring solutions. However, the possibility of taking quotients and connected sums implies that there are infinitely many topologies consistent with this theorem. For $D>6$, there is no simple description of the possible topologies.

The next step in the $D=4$ uniqueness proof for non-static black holes is Hawking's rigidity theorem \cite{Hawking:1973uf}, the statement that a stationary, rotating, analytic, black hole solution must be axisymmetric. This has been generalised to $D>4$: a stationary, rotating, analytic, black hole solution must admit a rotational symmetry (a $U(1)$ isometry acting as a rotation at infinity) \cite{Hollands:2006rj,Moncrief:2008mr}. However, this theorem guarantees only {\it one} rotational symmetry whereas the known explicit solutions have $N>1$ commuting rotational symmetries, i.e., more symmetry than guaranteed by the theorem. This suggests that there may exist other $D>4$ black hole solutions which have less symmetry than the known solutions. We will discuss this further in the next section.

Progress with classifying black holes can be made if one restricts attention to the case of multiple rotational symmetries. The classification of $D=5$ stationary rotating vacuum black holes with two rotational symmetries was studied in Ref. \cite{Hollands:2007aj} (extending \cite{Morisawa:2004tc}). It was shown that two such solutions are isometric if, and only if, they have the same mass, angular momenta, and {\it rod structure}. The latter (introduced in Refs. \cite{Emparan:2001wk,Harmark:2004rm}) encodes the angular velocity of the horizon and the nature of "axes of rotation" in the spacetime. The assumed symmetries imply that the horizon topology must be $S^3$, $S^1 \times S^2$ or a lens space (a quotient of $S^3$). It is not known whether (asymptotically flat) solutions of the latter kind exist. 

\subsection{Perturbative solutions}

Given the difficulty in finding explicit solutions of the Einstein equation, perturbative techniques have been used to obtain some insight into what other solutions might exist. Of particular interest are the possible topologies of higher-dimensional black holes, and the question of whether there exist higher-dimensional black holes with just one rotational symmetry. Two techniques have been used to investigate these questions.

The "blackfold" technique is a method for constructing black hole solutions whose horizons exhibit a large hierarchy of length scales. An example is a black ring for which the $S^1$ radius is much greater than the $S^{D-3}$ radius \cite{Emparan:2007wm}. Such large-radius black ring solutions are constructed perturbatively as an expansion in the ratio of these radii. More generally, the blackfold method has been used to construct approximate higher-dimensional solutions with topologies of the form $S^{p_1} \times S^{p_2} \times \ldots \times S^{p_K} \times s^{q}$ where 
$p_1, \ldots p_K$ are odd, $q \ge 2$, $S^{p_1}, \ldots S^{p_k}$ are large radius spheres and $s^q$ is a small radius sphere \cite{Emparan:2009cs}. 

The blackfold approach has also been used to construct solutions with just one rotational symmetry. Ref. \cite{Emparan:2009vd} presented a perturbative solution describing a $D=5$ "helical" black ring. This solutions can be visualised by imagining a spring formed into a loop. 

Another perturbative approach is to study linearized perturbations of an explicit solution. Non-uniform black strings provide a nice example of this. A uniform black string admits a time-independent linearized perturbation corresponding to the "threshold mode" of the Gregory-Laflamme instability. This time-independent perturbation exists precisely at the point where the non-uniform black string family bifurcates from the uniform string family. It corresponds to a non-uniform string solution with infinitesimal non-uniformity. Thus by studying linearized perturbations of the uniform string one can infer the existence of the non-uniform string family.

This method has been applied to perturbations of Myers-Perry black holes. Consider a singly spinning MP black hole with $D > 5$. As explained above, Ref. \cite{Dias:2009iu} showed that there is a critical $J_1$ (for given $M$) at which such a solution admits a non-trivial time-independent linearized perturbation, corresponding to the threshold mode of the ultraspinning instability. Earlier, Ref. \cite{Emparan:2003sy} had suggested that such a perturbation should be interpreted as evidence for a new family of black holes that bifurcates from the MP family. This new family would possess the same symmetries as the MP solution but with a slightly deformed horizon, corresponding to a depression at the poles of the sphere. It was suggested that moving along this new branch of solutions, the depression would increase, corresponding to an increasingly "pinched" sphere. Eventually, the sphere is expected to "pinch off" completely, and merge with the black ring family \cite{Emparan:2007wm}. 

The same strategy has been applied to perturbations of cohomogeneity-1 MP black holes (odd $D$ with $J_1 = J_2 = \ldots = J_N$). Recall that Refs. \cite{Dias:2010eu,Dias:2011jg} demonstrated an instability of such solutions near extremality. Again there is a time-independent threshold mode for this instability and so this was interpreted as evidence for a new branch of solutions bifurcating from the MP family. However, in this case, the threshold mode does not preserve the symmetries of the background geometry. In general it preserves only the rotational symmetry whose existence is guaranteed by the rigidity theorem. Hence the new family of solutions should have just one rotational symmetry. Note that these black holes would have spherical topology. Furthermore, in this case, there is not just one threshold mode but a multi-parameter set of them. If each of these extends to the new branch of solutions then these new solutions would have many more parameters than the MP family. For example, in $D=9$ they would have 70 parameters \cite{Dias:2010eu}, many more than the 5 parameters of the MP solution.

\section*{Acknowledgments}

I would like to thank Pau Figueras and Keiju Murata for permission to reproduce figures from Ref. \cite{Figueras:2012xj} and Roberto Emparan for permission to reproduce figures from Ref. \cite{Emparan:2008eg}. I am supported by a Royal Society University Research Fellowship and by European Research Council grant no. ERC-2011-StG 279363-HiDGR.

\section*{References}

\bibliography{reall_ae100}

\end{document}